# Development of a precise size-controllable pellet injector for the detailed studies of ablation phenomena and mechanism

K.Ichizono[1], S.Kugimiya[1], S.Nourgostar[1], K.N.Sato[2], TRIAM Exp.Group[2]
[1] Interdisciplinary Graduate School of Engineering Sciences, Kyushu University
[2] Research Institute for Applied Mechanics, Kyushu University

## 1. Introduction

From the viewpoint of performance of nuclear fusion plasmas, pellet injection experiments have been actively carried out in many toroidal devices in the sense of the control of density profile, obtaining high density or improved confinement, and diagnostic purposes. In order to have a common measure of pellet ablation, the regression study has been performed as an international cooperation activity, obtaining "IPAD" (International Pellet Ablation Database) [1]. However, these are empirical ones, and the mechanism of pellet ablation still remains to be studied.

According to such database or calculations based on the typical pellet ablation model (such as, so-called the neutral gas shielding model), it is known that the penetration depth into plasma is always quite sensitive to the pellet size. Also, an effective or suitable range of the pellet size for a certain plasma is generally very narrow, and this range largely varies depending to each toroidal plasma size. These typical characteristics are seen in the calculation results for the JIPP T-IIU tokamak plasma (Fig.2) and for the TEXTOR plasma (Fig.3).

Thus, the precise controllability of the pellet size, especially the size controllability with continuously variable system will be quite effective in order to carry out the detailed studies on pellet ablation and associated phenomena.

A pellet injector of new type with precisely and continuously controllable system of pellet size has started to be developed. This has a unique mechanics and structure of producing a frozen pellet in extremely low temperature region. The central part of the pellet injector with continuously size-variable system is given in Fig.5 (As a comparison, the conventional type is shown in Fig.4.). In the device presently developed in this research, we will precisely adjust the length of the cylindrical pellet ($\Phi$ 1.0mm) from 0.5 to 3mm by using the special "length limiting rod".

## 2. Numerical analysis by Neutral-Gas-Shielding (NGS) model

In pellet injection, in order to obtain optimum injection parameter (size, speed and angle etc.), it is necessary to calculate ablation rate, namely particle deposition for ice pellet. In this study the numerical analysis has been carried out by using the Neutral-gas-shielding model in order to estimate the effect of pellet size on the penetration depth and plasma parameters. In this ablation model, pellet (neutral) particles ablate with heat flux in plasma to form the ablation cloud around the pellet. Then, the cloud shields the pellet from energy flux from plasma. The ablation rate (regression speed of the pellet surface) is expressed as

$$\dot{r}_{pel} \propto r_{pel}^{-2/3} n_{e\infty}^{1/3} T_{e\infty}^{1.64}$$

where $r_{pel}$ is the pellet radius, $n_{e\infty}$ the electron density in plasma, and $T_{e\infty}$ the electron temperature in plasma.

Figure 1 shows that typical result of the ablation rate in JIPP T-II U plasma as a function of the pellet pass, where the pellet size is varied from 0.5 to 3.0 mm. In the area where size of the pellet is 1.0mm or less, the pellet completely ablated in plasma. On the other hand, in the area where size of the pellet is bigger than 1.0mm, the pellet passes the plasma without completely ablating. As is seen, the penetration depth is very sensitive to the pellet size. Figure 2 shows that typical results of the pellet penetration depth in JIPP T-II U plasma as a function of the pellet velocity, where the pellet size is varied from 0.1 to 1.0mm. Figure 3 shows that typical results of the pellet penetration depth in TEXTOR plasma as a function of the pellet velocity, where the pellet size is varied from 0.5 to 2.0 mm. As is seen, the penetration is quite sensitive to the pellet size, and also, an effective or suitable range of the pellet size for a certain plasma is generally very narrow, and this range largely varies depending on each toroidal plasma size.

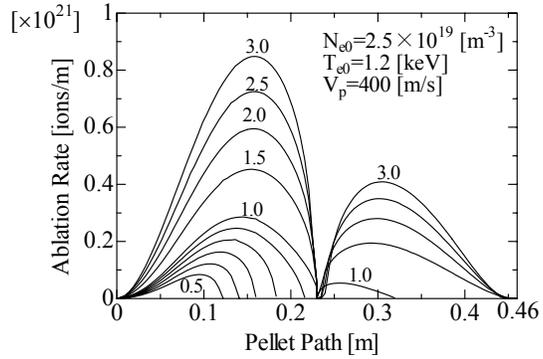

Figure 1. Typical of the ablation rate into JIPP T-II U plasma as a function of the pellet pass, where the pellet size is varied from 0.5 to 3.0mm.

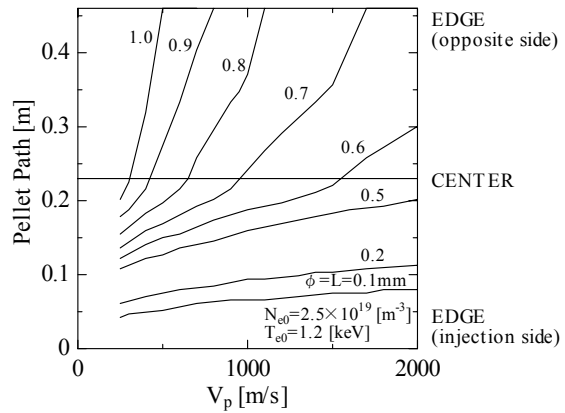

Figure 2. Typical results of the pellet penetration depth into JIPP TII-U plasma as a function of the pellet velocity, where the pellet size is varied from 0.1 to 1.0mm

## 3. Development of the pellet injector with continuously size-controllable system

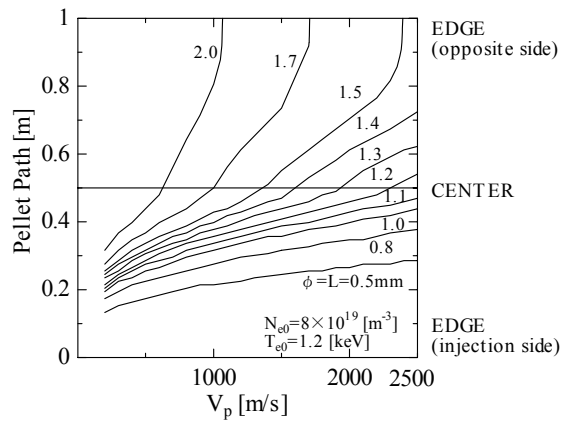

Figure 4 shows the central part of the pellet injector system of conventional type. In the system, since pellet size is decided by the thickness of the disk, one device can only make a pellet of one size. The central part of the continuously size-controllable pellet injection system being developed now is shown in Fig.5. In the device presently developed in this research, we will precisely adjust the length of the cylindrical pellet ($\Phi$ 1.0mm) from 0.5 to 3mm by using a special "length limiting rod".

Figure 3. Typical results of the pellet penetration depth into TEXTOR plasma as a function of the pellet velocity, where the pellet size is varied from 0.5 to 2.0mm

Detailed procedure of the pellet injector with continuously size-controllable system (in Fig.5) is shown below.

① Fill the hydrogen gas line (of part A) with hydrogen gas.
② Adjust pellet size by the length limiting rod (of part B).
③ Form a pellet by cooling (in part C).
④ Pull out the length limiting rod (of part B).
⑤ Rotate disk (of part D) by 180°.
⑥ Eject a pellet from propellant gas line (of part E).

The mechanism of the injector with " length limiting rod " has been designed and currently it is being tested.

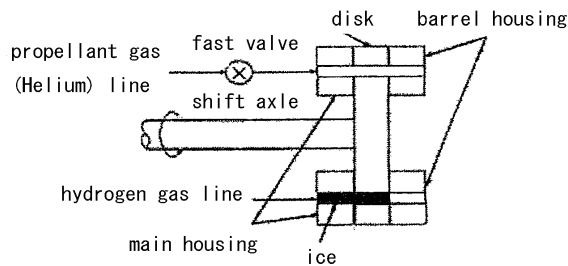

Figure 4. Central part of the pellet injector system of conventional type.

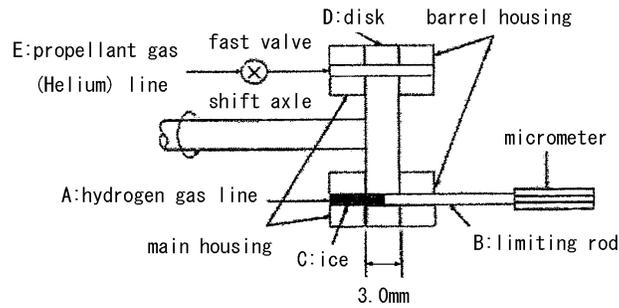

Figure 5. Central part of the continuously size-controllable pellet injection system.

### 4. Further steps in the development

The pellet injector system of this research is being developed in order to apply to the LHD research in the future. Therefore, the following issues are quite important to be established: ①Long term controlling, ②Remote operation, ③Total stability of the system. Thus, we have changed the system from the conventional liquid helium cooled type to the freezer type which might have caused an unfavorable temperature distribution. The following points should be considered: ①Precise mechanical drive controllability under very low temperature conditions. ②Periodic behavior of pellet production with a proper temperature distribution. ③Solving the adhering problem between pellet (solid hydrogen) and inner metallic surface of length limiting rod. ④Reproducibility of pellet cross section during the procedure of cutting by disk. ⑤ Reproducibility and stability of the whole cooling system.

### 5. Summary

A pellet injector system of new type with precisely and continuously controllable of pellet size has started to be developed. In the device presently developed in this research, we will precisely vary the length of the cylindrical pellet ($\Phi$ 1.0mm) from 0.5 to 3mm by using the special "length limiting rod".